\documentclass[twocolumn,english,showpacs,preprintnumbers,prb]{revtex4-1}
\usepackage[latin9]{inputenc}
\usepackage{color}
\usepackage{amsmath}
\usepackage{amssymb}
\usepackage{graphicx}
\usepackage{esint}

\makeatletter

\@ifundefined{textcolor}{}{%
 \definecolor{BLACK}{gray}{0}
 \definecolor{WHITE}{gray}{1}
 \definecolor{RED}{rgb}{1,0,0}
 \definecolor{GREEN}{rgb}{0,1,0}
 \definecolor{BLUE}{rgb}{0,0,1}
 \definecolor{CYAN}{cmyk}{1,0,0,0}
 \definecolor{MAGENTA}{cmyk}{0,1,0,0}
 \definecolor{YELLOW}{cmyk}{0,0,1,0}
}

\usepackage{color}

\usepackage{ulem}

\usepackage{aecompl}

\usepackage{epsfig}\usepackage{dcolumn}\usepackage{bm}

\usepackage{babel}

\makeatother

\usepackage{babel}
\begin{document}

\title{Decay of bound states in the continuum of Majorana fermions induced
by vacuum fluctuations: Proposal of qubit technology}

\author{L. S. Ricco$^{1}$, Y. Marques$^{1}$, F. A. Dessotti$^{1}$, R.
S. Machado$^{1}$, M. de Souza$^{2,}$}

\altaffiliation{Current address: Institute of Semiconductor and Solid State Physics, Johannes Kepler University Linz, Austria.}

\author{A. C. Seridonio$^{1,2}$}

\affiliation{$^{1}$Departamento de F\'{i}sica e Qu\'{i}mica, Unesp - Univ Estadual
Paulista, 15385-000, Ilha Solteira, SP, Brazil\\
 $^{2}$IGCE, Unesp - Univ Estadual Paulista, Departamento de F\'{i}sica,
13506-900, Rio Claro, SP, Brazil}
\begin{abstract}
We report on a theoretical investigation of the interplay between
vacuum fluctuations, Majorana quasiparticles (MQPs) and bound states
in the continuum (BICs) by proposing a new venue for qubit storage.
BICs emerge due to quantum interference processes as the Fano effect
and, since such a mechanism is unbalanced, these states decay as regular
into the continuum. Such fingerprints identify BICs in graphene as
we have discussed in detail in Phys. Rev. B \textbf{92}, 245107 and
045409 (2015). Here by considering two semi-infinite Kitaev chains
within the topological phase, coupled to a quantum dot (QD) hybridized
with leads, we show the emergence of a novel type of BICs, in which
MQPs are trapped. As the MQPs of these chains far apart build a delocalized
fermion and qubit, we identify that the decay of these BICs is not
connected to Fano and it occurs when finite fluctuations are observed
in the vacuum composed by electron pairs for this qubit. From the
experimental point of view, we also show that vacuum fluctuations
can be induced just by changing the chain-dot couplings from symmetric
to asymmetric. Hence, we show how to perform the qubit storage within
two delocalized BICs of MQPs and to access it when the vacuum fluctuates
by means of a complete controllable way in quantum transport experiments.
\end{abstract}


\maketitle

\section{Introduction}

An astonishing aftermath in the underlying framework of quantum theory
is the possibility of fluctuations within the corresponding quantum
field describing the vacuum, in which pairs of virtual particles pop
up leading to counterintuitive phenomena. In this regard, the Casimir
effect \cite{Casimir} is the most known picture in Physics arising
from the straight outcome of vacuum fluctuations. In particular, the
Casimir effect manifests itself as an attractive force between two
reflecting, plane and parallel plates, even when external fields are
entirely absent.

\begin{figure}[!]
\includegraphics[width=0.48\textwidth,height=0.24\textheight]{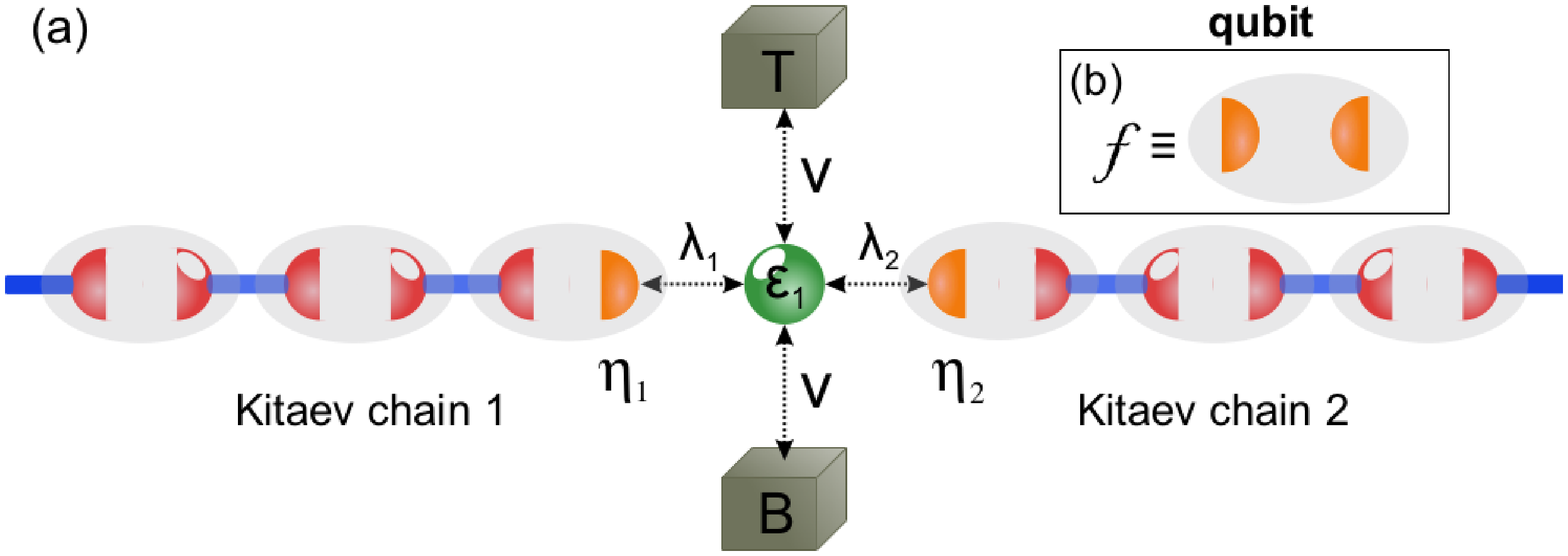}
\protect\protect\protect\protect\protect\protect\protect\protect\protect\protect\protect\protect\caption{\label{fig:Pic1} (Color online) For a off-resonance QD with the Fermi
level of the top (T) and bottom (B) leads, together with couplings
$\lambda_{1}=\lambda_{2},$ BICs of the MQPs $\eta_{1}$ and $\eta_{2}$
emerge. $V$ is the hybridization of the QD with the leads. The vacuum
of electron pairs for the quibt $f$ fluctuates when $\lambda_{1}\equiv(t+\Delta)\protect\neq\lambda_{2}\equiv(t-\Delta),$
thus inducing the decay of the BICs into the system continuum of energies.}
\end{figure}

On the ground of condensed matter Physics, we make explicit that the
interplay between vacuum fluctuations and Majorana quasiparticles
(MQPs) \cite{Franz} is accomplishable by the setup proposed in Fig.\,\ref{fig:Pic1},
where we find peculiar bound states in the continuum (BICs) \cite{Neuman}
for a pair of semi-infinite Kitaev chains \cite{Franz,UFMG,Nayana,Jelena,Vernek1,Vernek2}
in the topological phase and coupled to a quantum dot (QD) connected
to leads. Concerning on BICs, they were pioneering predicted by von
Neumann and Wigner in 1929 \cite{Neuman} as quantum states for electrons
described by localized square-integrable wave functions appearing
in the continuum of those delocalized and exhibiting infinite lifetimes
for such electrons. Hence, electrons within BICs do not decay into
the continuum acting as fully invisible states from the perspective
of conductance measurements. The issue on BICs had
a revival after the works of Stillinger and Herrick in 1975 \cite{SH},
followed by the experimental realization made by Capasso and co-workers in 1992,
concerning semiconductor heterostructures \cite{Capasso}.
Noteworthy, BICs are expected to emerge in several systems as in graphene
\cite{QPT,BIC1,Gong}, optics and photonics \cite{BIC2,BIC3,Exp1,Exp2},
setups characterized by singular chirality \cite{Chiral}, Floquet-Hubbard
states due to strong oscillating electric field \cite{Floquet} and
driven by A.C. fields \cite{AC}.

In this paper, we show that the setup proposed in Fig.\,\ref{fig:Pic1}
enables the observation of an unprecedented phenomenon: vacuum fluctuations
yielding the decay of BICs into the continuum, in which the building
blocks of the former are MQPs for qubit storage. We highlight two
setups considered proper platforms for MQPs in
Kitaev chains as well as for the experimental achievement of our proposal:
i) an $s$-wave superconductor nearby a semiconducting nanowire where
two magnetic fields exist perpendicular to each other, wherein one
of them arises from the spin-orbit coupling of the semiconductor,
while the second is applied externally to freeze the spin degree of
freedom of the system and to ensure topological superconductivity
\cite{wire1}, and ii) magnetic chains on top of superconductors characterized
by a strong spin-orbit parameter \cite{wire2,Jelena2}. Moreover,
MQPs are expected to rise among several setups as the fractional quantum
Hall state with filling factor $\nu=5/2$ \cite{QH}, in three-dimensional
topological insulators \cite{TI} and at the center of superconducting
vortices as well \cite{V1,V2,V3}.

To the best of our knowledge, the works up to date published in the
literature have focused mainly on BICs assisted by Fano interference
\cite{QPT,BIC1,BIC2,BIC3}. The so-called Fano effect is a quantum
inference phenomenon, wherein transport channels compete for the electron
tunneling, mainly via a continuum of energies hybridized with discrete
levels of nanoscale structures \cite{Fano1,Fano}. Here, as an alternative
we propose that quantum fluctuations in the vacuum of electron pairs
arising from the regular fermion and qubit $f$ composed by the MQPs
$\eta_{1}=\eta_{1}^{\dagger}$ and $\eta_{2}=\eta_{2}^{\dagger}$,
cf.\,shown in Fig.\ref{fig:Pic1}, give rise to the decay of these
peculiar BICs, the so-called quasi BICs. Otherwise, the BICs of MQPs
remain intact.

In order to present our proposal in a comprehensive way, we begin
by defining the qubit $f$ as follows: $f=\frac{1}{\sqrt{2}}(\eta_{1}+i\eta_{2})$
and $f^{\dagger}=\frac{1}{\sqrt{2}}(\eta_{1}-i\eta_{2}),$ in
which the occupations

\begin{align}
 & <f^{\dagger}f^{\dagger}>=\int_{-\infty}^{+\infty}d\varepsilon\text{{DO\ensuremath{S_{f^{\dagger}f^{\dagger}}}}}=<ff>=\int_{-\infty}^{+\infty}d\varepsilon\text{{DO\ensuremath{S_{ff}}}}\nonumber \\
 & =0\label{eq:number}
\end{align}
can be found \cite{book}, here expressed in terms of the densities
$\text{{DO\ensuremath{S_{f^{\dagger}f^{\dagger}}}}}$ and $\text{{DO\ensuremath{S_{ff}}},}$
since the pairings $f^{\dagger}f^{\dagger}$ and $ff$ are not allowed
in the system when both the Kitaev chains considered are equally coupled
to the QD. Later on, such densities will be deduced from our model
Hamiltonian. Furthermore, we will clarify that vacuum fluctuations
can be tunable experimentally. To that end, we should take into account
asymmetric Kitaev chain-dot couplings and a off-resonance QD with
the Fermi level of the leads, since the symmetric case prevents vacuum
fluctuations thus ensuring the qubit storage as MQPs delocalized at
the edges of the Kitaev chains as sketched in Fig.\,\ref{fig:Pic1}.
Hence, by means of BICs of MQPs, we propose a novel manner of qubit
storage when a single QD and controllable vacuum fluctuations are
accounted.

\section{The Model}

To give a theoretical description of the setup depicted in Fig.\,\ref{fig:Pic1}
describing two semi-infinite Kitaev chains within the topological
phase and connected to a QD coupled to leads, we employ an extension of
the Hamiltonian inspired on the original proposal from Liu and Baranger,
which is a spinless model to ensure topological superconductivity
Ref.\,{[}\onlinecite{Baranger}{]}:

\begin{eqnarray}
\mathcal{H} & = & \underset{\alpha k}{\sum}\tilde{\varepsilon}_{\alpha k}c_{\alpha k}^{\dagger}c_{\alpha k}+\varepsilon_{1}d_{1}^{\dagger}d_{1}+V\underset{\alpha k}{\sum}(c_{\alpha k}^{\dagger}d_{1}+\text{{H.c.}})\nonumber \\
 & + & \frac{(t+\Delta)}{\sqrt{2}}(d_{1}-d_{1}^{\dagger})\eta_{1}+i\frac{(\Delta-t)}{\sqrt{2}}(d_{1}+d_{1}^{\dagger})\eta_{2},\label{eq:TIAM}
\end{eqnarray}
where the electrons in the lead $\alpha=T,B$ are described by the
operator $c_{\alpha k}^{\dagger}$ ($c_{\alpha k}$) for the creation
(annihilation) of an electron in a quantum state labeled by the wave
number $k$ and energy $\tilde{\varepsilon}_{\alpha k}=\varepsilon_{k}-\mu_{\alpha}$,
with $\mu_{\alpha}$ as the chemical potential. For the QD coupled
to leads, $d_{1}^{\dagger}$ ($d_{1}$) creates (annihilates) an electron
in the state $\varepsilon_{1}.$ $V$ stands for the hybridizations
between the QD and the leads. The QD couples asymmetrically to the
Kitaev chains with tunneling amplitudes proportional to $(t+\Delta)\equiv\lambda_{1}$
and $(t-\Delta)\equiv\lambda_{2},$ respectively for the left and
right MQPs $\eta_{1}$ and $\eta_{2}.$ We stress that the prefactors
$1/\sqrt{2}$ and $i/\sqrt{2},$ respectively for $\lambda_{1}$ and
$\lambda_{2}$ constitute a convenient gauge that changes the last
two terms of Eq.\,(\ref{eq:TIAM}) into $td_{1}f^{\dagger}-td_{1}^{\dagger}f+\Delta f^{\dagger}d_{1}^{\dagger}-\Delta fd_{1}=td_{1}f^{\dagger}+\Delta f^{\dagger}d_{1}^{\dagger}+\text{{H.c.}},$
when the representation $f$ is adopted. As a result, we can notice
that the electrons within $f$ and $d_{1}$ beyond the normal tunneling
$t$ between them, become bounded as a Cooper pair with binding energy
$\Delta.$ Particularly, with $\Delta\neq0$ we will verify that the BICs here proposed decay into the continuum due to the emergence of these paring terms.

In what follows, we use the Landauer-Büttiker formula for the zero-bias
conductance $G$ \cite{Baranger}. Such a quantity is given by:
\begin{equation}
G=\frac{e^{2}}{h}\Gamma\int d\varepsilon\left(\frac{\partial f_{F}}{\partial\varepsilon}\right)\text{{Im}}(\tilde{\mathcal{G}}_{d_{1}^{\dagger}d_{1}}),\label{eq:G}
\end{equation}
where $\Gamma=2\pi V^{2}\sum_{k}\delta(\varepsilon-\varepsilon_{k})$
is the Anderson broadening \cite{Anderson}, $f_{F}$ stands for the
Fermi-Dirac distribution, $\tilde{\mathcal{G}}_{d_{1}^{\dagger}d_{1}}$
is the retarded Green's function for the QD in energy domain $\varepsilon,$
obtained from the time Fourier transform of $\tilde{\mathcal{G}}_{\mathcal{B^{\dagger}\mathcal{A}}}=\int d\tau\mathcal{G}_{\mathcal{B^{\dagger}\mathcal{A}}}e^{\frac{i}{\hbar}(\varepsilon+i0^{+})\tau}.$
Furthermore, we introduced $\mathcal{T}=-\Gamma\text{{Im}}(\tilde{\mathcal{G}}_{d_{1}^{\dagger}d_{1}})$
as the transmittance through the QD. $\mathcal{G}_{\mathcal{B^{\dagger}\mathcal{A}}}=-\frac{i}{\hbar}\theta(\tau){\tt Tr}\{\varrho[\mathcal{A}(\tau),\mathcal{B}^{\dagger}(0)]_{+}\}$
corresponds to the Green's function in time domain $\tau,$ here expressed
in terms of the density matrix $\varrho$ for Eq.\,(\ref{eq:TIAM})
and the Heaviside function $\theta\left(\tau\right).$ From $\mathcal{G}_{\mathcal{B^{\dagger}\mathcal{A}}},$
it is possible to find the expectation value $<\mathcal{B^{\dagger}\mathcal{A}}>=\int d\varepsilon\text{{DO\ensuremath{S_{\mathcal{B^{\dagger}\mathcal{A}}}}}}$
by using $\text{{DO\ensuremath{S_{\mathcal{B^{\dagger}\mathcal{A}}}}}}=-\frac{1}{\pi}\text{{Im}}(\mathcal{G}_{\mathcal{B^{\dagger}\mathcal{A}}})$
as the corresponding density of states, similarly to Eq.\,(\ref{eq:number})
for the vacuum. Particularly for Eq.\,(\ref{eq:G}), we used $\mathcal{A=B}=d_{1}$
and to calculate $\tilde{\mathcal{G}}_{d_{1}^{\dagger}d_{1}}$ together
with other Green's functions, we should employ the equation-of-motion
(EOM) method \cite{book} summarized as follows: $\omega\tilde{\mathcal{G}}_{B^{\dagger}\mathcal{A}}=(\varepsilon+i0^{+})\tilde{\mathcal{G}}_{B^{\dagger}\mathcal{A}}=[\mathcal{A},\mathcal{B^{\dagger}}]_{+}+\tilde{\mathcal{G}}_{\mathcal{B}^{\dagger}\left[\mathcal{A},\mathcal{\mathcal{H}}\right]}.$
As a result, we find

\begin{equation}
(\varepsilon-\varepsilon_{1}+i\Gamma)\tilde{\mathcal{G}}_{d_{1}^{\dagger}d_{1}}=1-t\mathcal{\tilde{G}}_{d_{1}^{\dagger},f}-\Delta\mathcal{\tilde{G}}_{d_{1}^{\dagger},f^{\dagger}},\label{eq:GF}
\end{equation}
in addition to the Green's functions $\mathcal{\tilde{G}}_{d_{1}^{\dagger},f}$
and $\mathcal{\tilde{G}}_{d_{1}^{\dagger},f^{\dagger}}.$ According
to the EOM approach, we also have $\omega\mathcal{\tilde{G}}_{d_{1}^{\dagger},f}=(\Delta\mathcal{\tilde{G}}_{d_{1}^{\dagger},d_{1}^{\dagger}}-t\mathcal{\tilde{G}}_{d_{1}^{\dagger},d_{1}}),$
$\omega\mathcal{\tilde{G}}_{d_{1}^{\dagger},f^{\dagger}}=(t\mathcal{\tilde{G}}_{d_{1}^{\dagger},d_{1}^{\dagger}}-\Delta\mathcal{\tilde{G}}_{d_{1}^{\dagger},d_{1}})$
and $\mathcal{\tilde{G}}_{d_{1}^{\dagger},d_{1}^{\dagger}}=-2t\Delta\tilde{K}\mathcal{\tilde{G}}_{d_{1}^{\dagger},d_{1}},$
in which $\tilde{K}=[\varepsilon+\varepsilon_{1}-K(t,\Delta)+i\Gamma]^{-1}K,$
$K(t,\Delta)=[\varepsilon^{2}+2i\varepsilon0^{+}-(0^{+})^{2}]^{-1}\omega(t^{2}+\Delta^{2})$
and $K=[\varepsilon^{2}+2i\varepsilon0^{+}-(0^{+})^{2}]^{-1}\omega.$
Consequently, the Green's function of the QD reads

\begin{align}
\tilde{\mathcal{G}}_{d_{1}^{\dagger}d_{1}} & =\frac{1}{\varepsilon-\varepsilon_{1}+i\Gamma-\Sigma_{\text{{MQPs}}}},\label{eq:d1d1}
\end{align}
where $\Sigma_{\text{{MQPs}}}=K(t,\Delta)+(2t\Delta)^{2}K\tilde{K}$
accounts for the self-energy due to the MQPs connected to the QD and
$\text{{DOS}}_{11}=-\frac{1}{\pi}\text{{Im}}(\tilde{\mathcal{G}}_{d_{1}^{\dagger}d_{1}})$
is the density of states for the QD. Particularly for $t=\Delta=\frac{\lambda}{\sqrt{2}},$
the expressions for $\tilde{K}$ and $\Sigma_{\text{{MQPs}}}$ found
in Ref.\,{[}\onlinecite{Baranger}{]} are recovered.

To perceive the emergence of BICs in the Kitaev chains and vacuum
fluctuations, we need to find the densities for the MQPs $\eta_{1}$
and $\eta_{2},$ namely $\text{{DOS}}_{\eta_{1}}=-\frac{1}{\pi}\text{{Im}}(\tilde{\mathcal{G}}_{\eta_{1}\eta_{1}})$
and $\text{{DOS}}_{\eta_{2}}=-\frac{1}{\pi}\text{{Im}}(\tilde{\mathcal{G}}_{\eta_{2}\eta_{2}}),$
together with $\text{{DOS}}_{ff}=-\frac{1}{\pi}\text{{Im}}(\tilde{\mathcal{G}}_{ff})$
and $\text{{DOS}}_{f^{\dagger}f^{\dagger}}=-\frac{1}{\pi}\text{{Im}}(\tilde{\mathcal{G}}_{f^{\dagger}f^{\dagger}}),$
in which the latter allows to determine the occupations $<f^{\dagger}f^{\dagger}>$
and $<ff>$ as Eq.\,(\ref{eq:number}) ensures for the vacuum of
electron pairs. Thus the EOM gives rise to

\begin{eqnarray}
\mathcal{\tilde{G}}_{\eta_{1}\eta_{1}} & = & \frac{1}{2}(\mathcal{\tilde{G}}_{f^{\dagger}f^{\dagger}}+\mathcal{\tilde{G}}_{ff^{\dagger}}+\mathcal{\tilde{G}}_{f^{\dagger}f}+\mathcal{\tilde{G}}_{ff})\label{eq:eta1}
\end{eqnarray}
and

\begin{eqnarray}
\mathcal{\tilde{G}}_{\eta_{2}\eta_{2}} & = & \frac{1}{2}(-\mathcal{\tilde{G}}_{f^{\dagger}f^{\dagger}}+\mathcal{\tilde{G}}_{ff^{\dagger}}+\mathcal{\tilde{G}}_{f^{\dagger}f}-\mathcal{\tilde{G}}_{ff})\label{eq:eta2}
\end{eqnarray}
for the Green's functions of the MQPs, with

\begin{equation}
\omega\tilde{\mathcal{G}}_{f^{\dagger}f^{\dagger}}=(t\tilde{\mathcal{G}}_{f^{\dagger}d_{1}^{\dagger}}-\Delta\mathcal{G}_{f^{\dagger}d_{1}})\label{eq:GFfdfd}
\end{equation}
and

\begin{equation}
\omega\tilde{\mathcal{G}}_{ff}=(-t\tilde{\mathcal{G}}_{fd_{1}}+\Delta\tilde{\mathcal{G}}_{fd_{1}^{\dagger}})\label{eq:GFff}
\end{equation}
for those describing the aforementioned vacuum. To close the system
of Green's functions above-described, we calculate via EOM the following
$\omega\tilde{\mathcal{G}}_{ff^{\dagger}}=(1+t\tilde{\mathcal{G}}_{fd_{1}^{\dagger}}-\Delta\mathcal{G}_{fd_{1}}),$
$\omega\tilde{\mathcal{G}}_{f^{\dagger}f}=(1-t\tilde{\mathcal{G}}_{f^{\dagger}d_{1}}+\Delta\tilde{\mathcal{G}}_{f^{\dagger}d_{1}^{\dagger}}),$
$\omega\tilde{\mathcal{G}}_{f^{\dagger}d_{1}}=-t(1+2\Delta^{2}\tilde{K})\tilde{\mathcal{G}}_{d_{1}^{\dagger}d_{1}},$
$\omega\tilde{\mathcal{G}}_{fd_{1}}=-\Delta(1+2t^{2}\tilde{K})\tilde{\mathcal{G}}_{d_{1}^{\dagger}d_{1}},$
$\tilde{\mathcal{G}}_{f^{\dagger}d_{1}^{\dagger}}=\Delta\tilde{K}K^{-1}\omega^{-1}-2t\Delta\tilde{K}\mathcal{G}_{f^{\dagger}d_{1}}$
and $\tilde{\mathcal{G}}_{fd_{1}^{\dagger}}=t\tilde{K}K^{-1}\omega^{-1}-2t\Delta\tilde{K}\mathcal{G}_{fd_{1}}.$

Thus based on the theoretical framework developed up to here, shortly
thereafter we will discuss the role of Eqs.(\ref{eq:d1d1}), (\ref{eq:eta1})
and (\ref{eq:eta2}) in the connection with the novel qubit technology
proposed in this paper.

\section{Results and Discussion}

\begin{figure}
\includegraphics[width=0.43\textwidth,height=0.22\textheight]{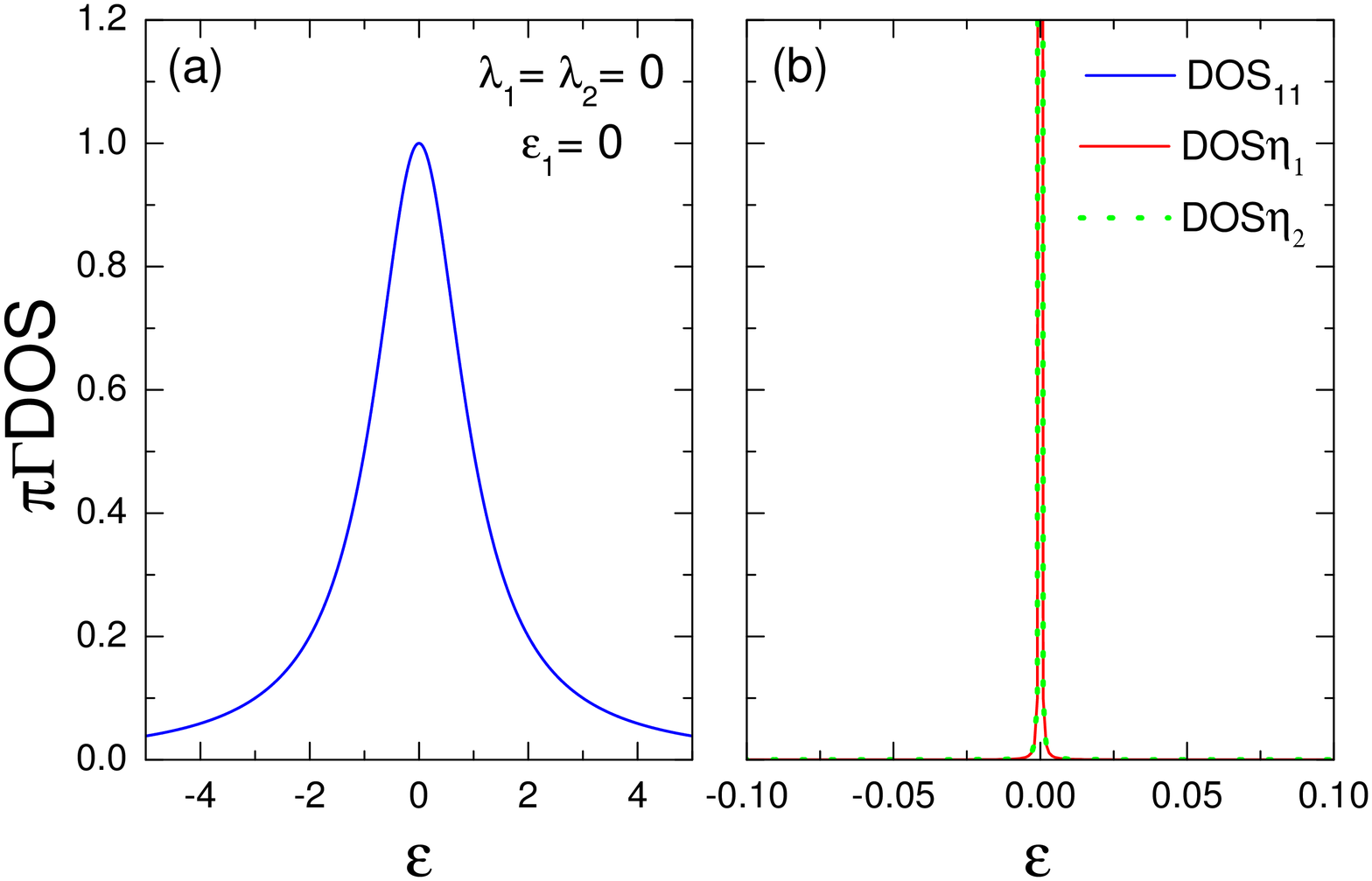}
\protect\protect\protect\protect\protect\protect\protect\protect\protect\protect\protect\protect\caption{\label{fig:Pic2}(Color online) (a) DOS of the QD in resonance with
the Fermi level of the leads $(\varepsilon_{F}\equiv0)$ when the
Kitaev chains are absent: a Lorentzian shape is observed. (b) DOSs
of the MQPs by neglecting the couplings to the leads: a Delta function
profile appears instead. Details in the main text.}
\end{figure}

In the simulations discussed here we adopt $T=0$ and the Anderson
broadening $\Gamma=2\pi V^{2}\sum_{k}\delta(\varepsilon-\varepsilon_{k})$
as the energy scale for the parameters from the system Hamiltonian
of Eq.\,(\ref{eq:TIAM}). In order to make explicit the phenomenon
of qubit storage ruled by vacuum fluctuations, due to the Kitaev chains
connected to the QD, we should begin by analyzing the cases in which
both are decoupled from each other $(\lambda_{1}=\lambda_{2}=0)$.
Within this situation, but for the QD in resonance with the Fermi
level of the metallic leads $(\varepsilon_{1}=\varepsilon_{F}\equiv0),$
the standard Lorentzian shape depicted in Fig.\,\ref{fig:Pic2}(a)
for the DOS encoded by $\text{{DOS}}_{11}=-\frac{1}{\pi}\text{{Im}}(\tilde{\mathcal{G}}_{d_{1}^{\dagger}d_{1}})$
as a function of energy $\varepsilon$ is verified. In the panel (b)
of the same figure, we find coincident profiles for $\text{{DOS}}_{\eta_{1}}=-\frac{1}{\pi}\text{{Im}}(\tilde{\mathcal{G}}_{\eta_{1}\eta_{1}})$
and $\text{{DOS}}_{\eta_{2}}=-\frac{1}{\pi}\text{{Im}}(\tilde{\mathcal{G}}_{\eta_{2}\eta_{2}})$
describing the DOSs of the MQPs, respectively found at the edges of
the Kitaev chains $1$ and $2$ nearby the QD. Once the MQPs are zero-energy
modes for discrete states, the curves of $\text{{DOS}}_{\eta_{1}}$
and $\text{{DOS}}_{\eta_{2}}$ are indeed Dirac delta functions as
expected, due to the absence of leads connected to the Kitaev chains.
These Delta functions represent the complete storage of the qubit
$f$ composed by the MQPs $\eta_{1}$ and $\eta_{2},$ since the zero
broadening of such DOSs point out that the electron within $f$ has
an infinite lifetime and does not decay into the QD. Below, we will
see that such a scenario is modified when the Kitaev chain-QD couplings
are turned-on, i.e., $\lambda_{1}=\lambda_{2}\neq0$.

\begin{figure}
\includegraphics[width=0.5\textwidth,height=0.27\textheight]{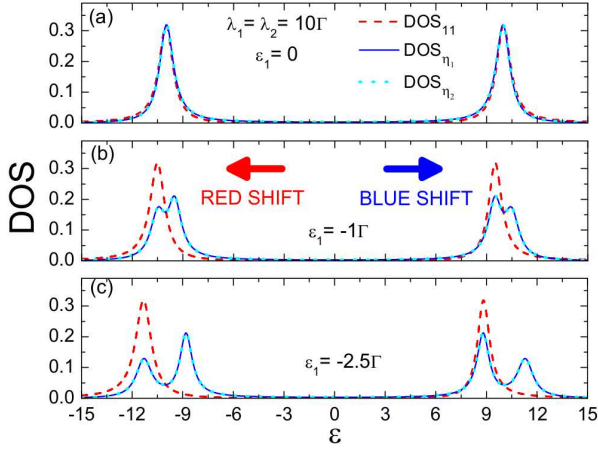}
\protect\protect\protect\protect\protect\protect\protect\protect\protect\protect\protect\protect\caption{\label{fig:Pic3}(Color online) (a) Scenario of Figs.\ref{fig:Pic2}(a)
and (b) modified when the Kitaev chain-QD symmetric couplings are
turned-on: as the QD and the MQPs are in resonance, their DOSs split
equally. (b) For the QD off-resonance, a new structure composed by
four peaks emerges only in the DOSs for the MQPs. (c) Visualization
of the red and blue shifts of the peaks found in (b) by placing the
QD very far from the resonance. }
\end{figure}

\begin{figure}
\includegraphics[width=0.45\textwidth,height=0.27\textheight]{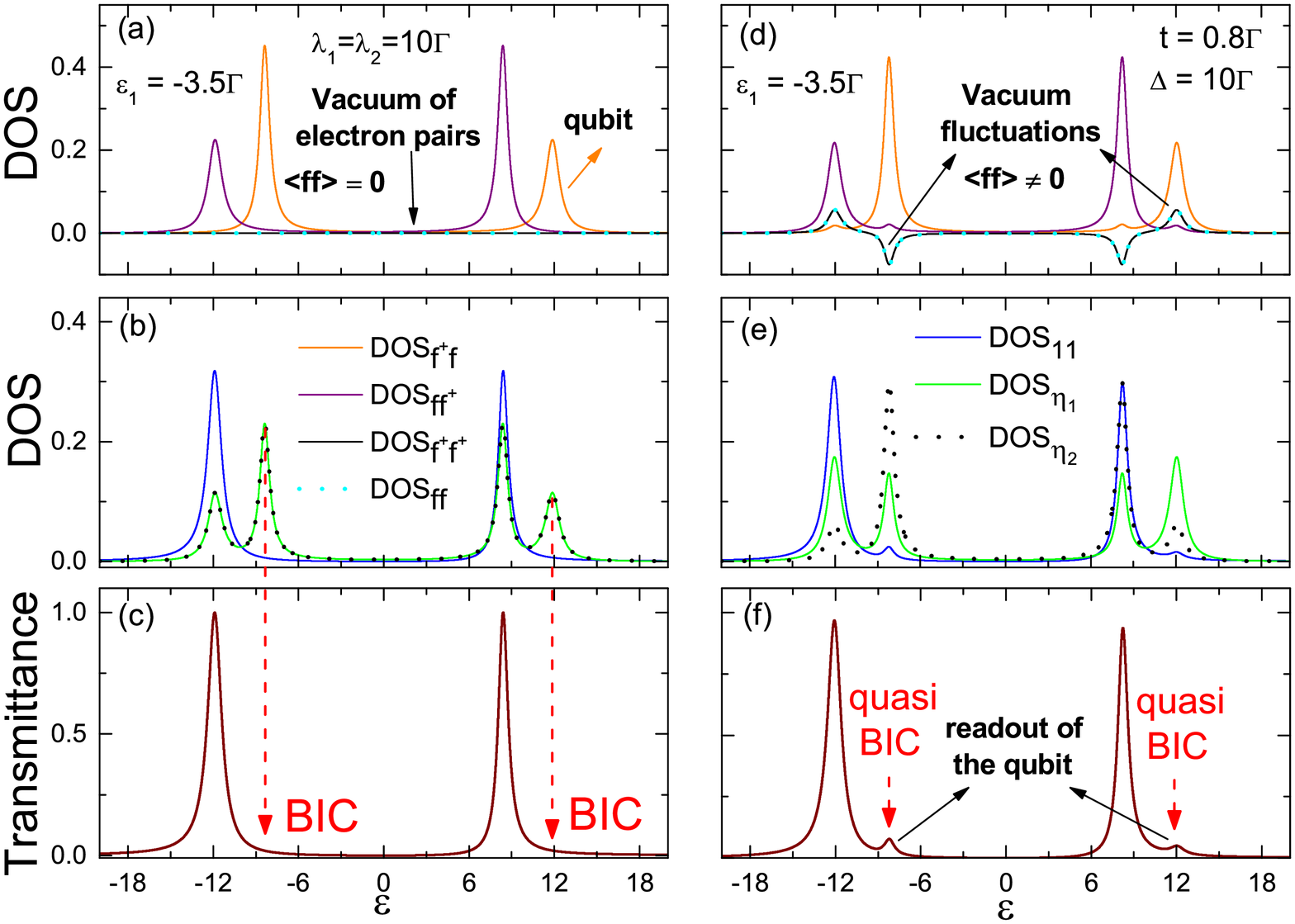}
\protect\protect\protect\protect\protect\protect\protect\protect\protect\protect\protect\protect\caption{\label{fig:Pic4}(Color online) Symmetric regime of Kitaev chain-QD
couplings: (a) DOSs for electrons within the qubit $f$ and absence
of vacuum fluctuations outlined by the vertical arrow; (b) DOSs for
the QD and MQPs in absence of vacuum fluctuations; (c) transmittance
profile with BICs appearing denoted by dashed-vertical arrows, i.e.,
the lacking of the corresponding peaks found at the same positions
within panel (b) for the DOSs of the MQPs is a BIC fingerprint, since
these states do not contribute to the quantum transport. Only the
peaks of the DOS for the QD remain in the transmittance. Asymmetric
regime of Kitaev chain-QD couplings: (d) the vacuum fluctuates around
the BICs verified in (c) and the DOSs for $f$ also change; (e) DOSs
for the QD and MQPs in presence of vacuum fluctuations; (f) As aftermath
of the fluctuations observed in (d) which are denoted by arrows, the
BICs of MQPs found in panel (c) decay as quasi BICs. They are found
outlined by dashed-vertical arrows as we can visualize in the transmittance
profile.}
\end{figure}

Fig.\,\ref{fig:Pic3}(a) treats the symmetric regime $\lambda_{1}=\lambda_{2}=10\Gamma,$
in which the QD is still in resonance with the leads Fermi level.
As both the QD and MQPs are zero-energy modes and share the same DOS
profile $(\text{{DOS}}_{\eta_{1}}=\text{{DOS}}_{\eta_{2}}=\text{{DOS}}_{11}),$
the outcome of this set is to exhibit the same splitting of the zero-peak,
which was originally centered at the Fermi energy as showed in Figs.\ref{fig:Pic2}(a)
and (b). Once the zero-peak is splitted, one can propose a manner
of controlling this splitting within the $\text{{DOS}}_{11}.$ The
way we have found is by placing the QD off-resonance in respect with
the leads Fermi level. This picture can be visualized in Fig.\,\ref{fig:Pic3}(b)
for the dashed-red curve with $\varepsilon_{1}=-\Gamma,$ where we
can perceive the red and blue shifts of the peaks. On the other hand,
the response of the MQPs due to the tuning of $\varepsilon_{1}$ is
fully different compared to the QD: the original pattern given by
a pair of peaks for the MQPs appearing in Fig.\,\ref{fig:Pic3}(a)
evolves towards a novel structure, where four peaks emerge as depicted
by the dotted and solid blue curves of Fig.\,\ref{fig:Pic3}(b).
Note that this novel pattern is not well resolved for $\varepsilon_{1}=-\Gamma$
yet, since the peaks within each pair of peaks are found partially
merged. However, if we consider $\varepsilon_{1}=-2.5\Gamma$ as in
Fig.\,\ref{fig:Pic3}(c), the visibility of the four peaks becomes
more pronounced and they appear completely resolved. We should draw
attention in the manner that the pairs of peaks in the DOSs for the
MQPs evolve from the pattern observed in Fig.\,\ref{fig:Pic3}(b)
to that in panel (c). To reveal the underlying mechanism
of such an electron-hole asymmetry, we should focus on panels (a)-(c)
of Fig.\,\ref{fig:Pic4} and, in particular, Eqs.(\ref{eq:eta1})
and (\ref{eq:eta2}) for $\text{{DOS}}_{\eta_{1}}=-\frac{1}{\pi}\text{{Im}}(\tilde{\mathcal{G}}_{\eta_{1}\eta_{1}})$
and $\text{{DOS}}_{\eta_{2}}=-\frac{1}{\pi}\text{{Im}}(\tilde{\mathcal{G}}_{\eta_{2}\eta_{2}}),$
respectively.

\begin{figure}
\includegraphics[width=0.5\textwidth,height=0.27\textheight]{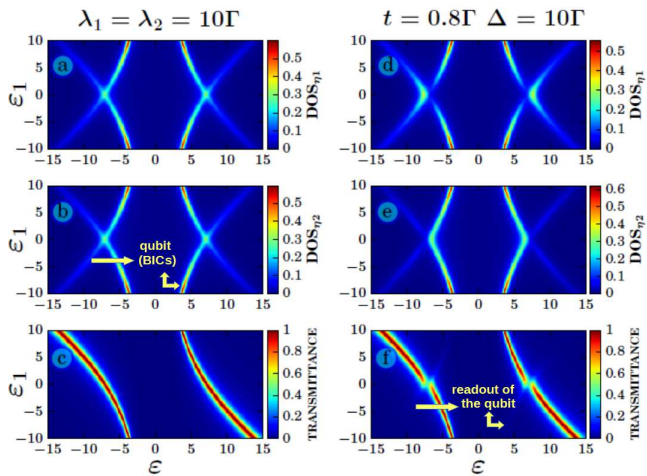}
\protect\protect\protect\protect\protect\protect\protect\protect\protect\protect\protect\protect\caption{\label{fig:Pic5}(Color online) Symmetric regime of Kitaev chain-QD
couplings: in panels (a), (b) and (c) we see the density plots of
the DOSs for the MQPs and transmittance spanned by the axis $\varepsilon_{1}$
and $\varepsilon.$ Panel (c) for the transmittance shows only two
bows from the set of four bows found in (a) and (b) describing the
DOSs of the MQPs. The lack of such bows in the transmittance then
denotes BICs of MQPs. Asymmetric regime of Kitaev chain-QD couplings:
panels (d) and (e) are similar to (a) and (b), but in panel (f) quasi
BICs appear as light bows in the transmittance in respect with those
verified in panels (d) and (e).}
\end{figure}

Still in the symmetric regime $\lambda_{1}=\lambda_{2}=10\Gamma,$
we see that only $\mathcal{\tilde{G}}_{ff^{\dagger}}$ and $\mathcal{\tilde{G}}_{f^{\dagger}f}$
rise in Fig.\,\ref{fig:Pic4}(a) respectively via the $\text{{DOS}}_{ff^{\dagger}}=-\frac{1}{\pi}\text{{Im}}(\tilde{\mathcal{G}}_{ff^{\dagger}})$
and $\text{{DOS}}_{f^{\dagger}f}=-\frac{1}{\pi}\text{{Im}}(\tilde{\mathcal{G}}_{f^{\dagger}f})$,
while $\mathcal{\tilde{G}}_{f^{\dagger}f^{\dagger}}$ and $\mathcal{\tilde{G}}_{ff}$
contributes with $\text{{DOS}}_{f^{\dagger}f^{\dagger}}=\text{{DOS}}_{ff}=0$
as aftermath of the vacuum $<f^{\dagger}f^{\dagger}>=<ff>=0$ for
the electron pairs. Besides, the structure of four peaks firstly
displayed in Fig.\,\ref{fig:Pic3}(c) and together with Fig.\,\ref{fig:Pic4}(b)
for the MQPs $\eta_{1}$ and $\eta_{2}$ considering $\varepsilon_{1}=-3.5\Gamma,$
then make explicit that the unmatched profiles of
the $\text{{DOS}}_{ff^{\dagger}}$ and $\text{{DOS}}_{f^{\dagger}f}$
play the role of two spectral functions analogous to the possibilities $\text{{DOS}}_{\uparrow}$
and $\text{{DOS}}_{\downarrow},$ due to spin-imbalance in ferromagnetic
systems. Based on this, we can realize the features within Fig.\,\ref{fig:Pic4}(c),
where the peaks appearing in the transmittance $\mathcal{T}=-\Gamma\text{{Im}}(\tilde{\mathcal{G}}_{d_{1}^{\dagger}d_{1}})$
determined by Eq.(\ref{eq:G}) do not correspond to those found in
panel (a) for the $\text{{DOS}}_{f^{\dagger}f},$ but just to those
from $\text{{DOS}}_{ff^{\dagger}}.$ Equivalently, from those four
peaks observed in Fig.\,\ref{fig:Pic4}(b) for the MQPs, solely two
of them decay into the QD, thus contributing to the conductance. As
a result, the peaks within $\text{{DOS}}_{\eta_{1}}=\text{{DOS}}_{\eta_{2}}$
that do not appear in $\mathcal{T}$ correspond to BICs of MQPs, which
constitute the building blocks of the delocalized qubit $f$ characterized
by the two peaks found in Fig.\,\ref{fig:Pic4}(a) for the $\text{{DOS}}_{f^{\dagger}f}.$
These invisible peaks arising from the $\text{{DOS}}_{f^{\dagger}f}$
in $\mathcal{T}$, then provide a novel manner of qubit storage, wherein
despite the coupling $\lambda_{1}=\lambda_{2}=10\Gamma$ of the Kitaev
chains with the QD, the decay of the state $f$ is completely prevented.
Such a forbiddance, we should highlight, occurs when
the vacuum $<f^{\dagger}f^{\dagger}>=<ff>=0$ does not fluctuate {[}Fig.\,\ref{fig:Pic4}(a){]}.
In what follows, we will show that when $<f^{\dagger}f^{\dagger}>=<ff>\neq0,$
just in the vicinity of BICs, the suppression of the qubit storage
occurs and hence, the decay of the BICs as quasi BICs is allowed appearing
thought out the transmittance.

To fluctuate the vacuum considered, it is demanded
asymmetric couplings $\lambda_{1}\neq\lambda_{2}$ by means of $\Delta=10\Gamma$
and $t=0.8\Gamma,$ for instance. As a net effect we have $\text{{DOS}}_{f^{\dagger}f^{\dagger}}=\text{{DOS}}_{ff}\neq0$
corresponding to fluctuations in the vacuum $<f^{\dagger}f^{\dagger}>=<ff>\neq0$
around the BICs, which appear pointed out by black arrows in Fig.\,\ref{fig:Pic4}(d).
In such a situation, pairing terms $d_{1}f+f^{\dagger}d_{1}^{\dagger}$
appear, which allow the correlations above to become finite. Consequently,
the profiles for $\text{{DOS}}_{\eta_{1}}$ and $\text{{DOS}}_{\eta_{2}}$
become distinct as showed in Fig.\,\ref{fig:Pic4}(e), resulting
in the detection of the BICs by means of the quasi BICs, which appear
as unpronounced states indicated by dashed-red arrows in Fig.\,\ref{fig:Pic4}(f).
Noteworthy, the quasi BICs are placed exactly at the positions of
the BICs found in Fig.\,\ref{fig:Pic4}(c). Moreover, it is worth
noticing in opposite to the unbalance of Fano interference as the
underlying mechanism for the rising of quasi BICs reported in graphene
systems \cite{QPT,BIC1}, here we identify that vacuum fluctuations
of the electron pairs described by the expectation value $<f^{\dagger}f^{\dagger}>=<ff>\neq0$
as the trigger for the decay of these peculiar BICs of MQPs. When
it occurs, the information within the qubit is read via transmittance.

To summarize the results presented up to here, we wrap up them in
the density plots of $\mathcal{T}$ spanned by the axis $\varepsilon_{1}$
and $\varepsilon$ appearing in Fig.\,\ref{fig:Pic5}, where panels
(a)-(c) and (d)-(e) designate respectively, the symmetric and asymmetric
regimes of couplings between the QD and the Kitaev chains. Panels
(a), (b), (d) and (e) of the same figure, in particular, share a main
characteristic: all of them exhibit the structure of four peaks previously
reported, which appear as four bows in the density plot format. As
just two bows from the set of four found in Figs.\ref{fig:Pic5}(a)
and (b) are displayed in (c), those absent are then BICs of MQPs,
while the light pair of bows in (f) represent quasi BICs.

\section{Conclusions}

In summary, we have proposed a setup based on two semi-infinite Kitaev
chains presenting MQPs at their edges both coupled to a single QD
crossed by a current due to source and drain reservoirs of electrons,
in which BICs of MQPs are revealed as building blocks for the storage
of a delocalized qubit. For absence of fluctuation in the vacuum of
electron pairs as aftermath of the delocalized fermion and qubit builded
by these MQPs, the BICs do not decay into the system continuum and
are still unperceived by conductance measurements, which then ensure
the storage. Fluctuations of the aforementioned vacuum then trigger
the decay of such states as quasi BICs. Distinct from the standard
BICs formed by Fano effect \cite{QPT,BIC1,BIC2,BIC3}, the corresponding
for MQPs are ruled by vacuum fluctuations, thus constituting a novel
phenomenon. Experimentally speaking, it can be feasible just by tuning
the Kitaev chain-dot couplings from symmetric (intact BICs where the
qubit is found) to asymmetric (vacuum fluctuations induced), when
the QD is off-resonance with the Fermi energy of the metallic leads.

\section{Acknowledgments}

This work was supported by CNPq, CAPES, 2014/14143-0 and 2015/23539-8 S{ã}o Paulo
Research Foundation (FAPESP).


\end{document}